\documentclass[10pt]{article}
\usepackage{amsfonts,amsmath,amssymb}
\usepackage{epsfig,comment}
\usepackage{xspace}

\usepackage[dvips=true,bookmarks=true]{hyperref}

\title{\textbf{Cooling algorithms based on the 3-bit majority}}
\author{Phillip Kaye}

\newcommand{\B}{\mathcal{B}}
\newcommand{\init}{\text{i}}
\newcommand{\steady}{\text{steady}}

\newcommand{\eps}{\varepsilon}

\newcommand{\ceil}[1]{\left\lceil #1 \right\rceil}

\newcommand{\CNOT}{\mbox{\sc cnot}\xspace}
\newcommand{\SWAP}{\mbox{\sc swap}\xspace}
\newcommand{\NOT}{\mbox{\sc not}\xspace}

\newcommand{\qed}{\hspace{5mm}\square\,\vspace{4mm}}
\renewcommand{\mod}[1]{\text{ mod } #1}

\renewcommand{\th}{\text{th}}
\renewcommand{\lim}{\text{lim}}

\newcommand{\TBC}{\textnormal{3BC}}
\newcommand{\hb}{\textnormal{hb}}

\newtheorem{definition}{Definition}

\newtheorem{lemma}{Lemma}

\newtheorem{claim}{Claim}

\numberwithin{fact}{section}

\begin{document}
\maketitle

\begin{abstract}
Algorithmic cooling is a potentially important technique for making
scalable NMR quantum computation feasible in practice.  Given the
constraints imposed by this approach to quantum computing, the most
likely cooling algorithms to be practicable are those based on
simple reversible polarization compression (RPC) operations acting
locally on small numbers of bits.  Several different algorithms
using 2- and 3-bit RPC operations have appeared in the literature,
and these are the algorithms I consider in this note.  Specifically,
I show that the RPC operation used in all these algorithms is
essentially a majority vote of 3 bits, and prove the optimality of
the best such algorithm. I go on to derive some theoretical bounds
on the performance of these algorithms under some specific
assumptions about errors.
\end{abstract}

\section{Background}

Consider a probabilistic bit that equals 0 with probability $p$.
Define the {\it bias} of the bit to be
\[\B=p-(1-p)=2p-1,\]
which is the difference between the probability that the bit equals
0 and the probability that the bit equals 1. (The symbol ``$\eps$''
is usually used to denote the bias in the literature on algorithmic
cooling; I prefer to reserve this symbol for error rates.) The
problem addressed by algorithmic cooling is the following. Given
some number of bits initially having a common bias $\B_\init > 0$,
distill out some smaller number of bits having greater bias. This
should be accomplished without the need for any pure ancillary bits
initialized to 0, since preparing such initialized bits is the
problem to be solved. Also, we should assume that we cannot perform
measurements.

Algorithmic cooling has significant relevance to quantum computing,
because for physical systems like nuclear spins controlled using
nuclear magnetic resonance (NMR), obtaining a pure initial state can
be very challenging. It is this fact that has motivated recent
research on the implementation of algorithmic cooling in NMR quantum
computers, as well as theoretical investigations of the efficiency
and performance of cooling algorithms.

Algorithmic cooling in the context of NMR quantum computation first
appeared in \cite{SV99}.  The authors presented a method for
implementing {\it reversible polarization compression} (RPC).  The
idea of RPC is to use reversible logic to implement a permutation on
the (classical) states of $n$ bits, so that the bias of some of the
bits is increased, while the bias of others is decreased (this is
closely connected to data compression).  Unfortunately RPC is
theoretically limited by Shannon's Bound, which says that the
entropy of a closed system cannot decrease.

An alternative algorithm was proposed in \cite{BMRVV} to enable
cooling below the Shannon bound. The idea is to use a second
register of bits that quickly {\it relax} to the initial bias
$\B_\init$. Call these the {\it relaxation bits}, and refer to the
bits on which we perform the RPC operation as the {\it compression
bits}. The idea is to first use RPC to increase the bias of some of
the compression bits, while decreasing the bias of the other
compression bits. Then the hotter compression bits (i.e. those
having decreased bias) are swapped with the relaxation bits, where
they will quickly relax back to the initial bias $\B_\init$.
Repeating this procedure effectively pumps heat out of the some of
the compression bits, cooling them to bias much higher than
$\B_\init$. This system is analogous to a kitchen refrigerator,
where the relaxation bits behave like the radiator on the back of
the refrigerator, dumping the heat taken from the refrigerator
compartment out into the surrounding environment.  This approach is
often referred to as ``heat-bath algorithmic cooling'', and the
relaxation bits are often referred to as the ``heat bath''.

Another approach to heat-bath algorithmic cooling was introduced in
\cite{FLMR04}.  Their algorithm has a simpler analysis than the
algorithm in \cite{BMRVV}, and gives a better bound on the size of
molecule required to cool a single bit.

In \cite{SMW05} the physical limits of heat-bath cooling are
explored. In their analysis, the assumption is that the basic
operations can be implemented perfectly, without errors. Even given
this assumption, the authors show that if the heat bath temperature
is above a certain temperature threshold, no cooling procedure can
initialize the system sufficiently for quantum computation.  A
heat-bath cooling algorithm called the ``partner pairing algorithm''
(PPA) is introduced to derive bounds on the best possible
performance of algorithmic cooling with a heat bath. The PPA
performs better than the previous algorithms, but it is unclear
whether implementing the required permutations will be realistic in
practice.  In this paper I will focus on cooling algorithms based on
repeated application of simple 2 or 3-bit RPC steps.

\section{Architecture}\label{sec_architecture}

To be useful for NMR quantum computing, we should implement cooling
algorithms on a register of quantum bits all having some initial
bias $\B_\init$, without access to any ``clean'' ancillary bits.
Further, we should be careful about how much ``local control'' we
assume is directly provided by the system.  In \cite{SV98}, four
primitive computational operations are proposed as being supported
by NMR quantum computers.  For implementing the cooling algorithm,
the first two of these suffice:
\begin{enumerate}
\item[$o_1$)] Cyclically shift the $n$ bits clockwise or
counterclockwise one position.
\item[$o_2$)] Apply an arbitrary two-bit operation to the first two bits (i.e. to the bits under a fixed ``tape-head'').
\end{enumerate}
To implement the two operations, \cite{SV98} suggested to use a
repeating polymer like the $ABC$-chains used for global control
schemes (e.g. \cite{Llo93}). The chain could be configured as a
closed loop.  To mark the position of the ``first two bits'' of the
chain (for operation $o_2$), an atom of a fourth type, $D$, is
positioned adjacent to the chain, in the desired location.

Notice that a system supporting operations $o_1$ and $o_2$ above can
be re-phrased in terms of a fixed ``tape'' containing the
bit-string, and a moving ``head'' that can be positioned over any
adjacent pair of tape cells.  For convenience, the tape can be
viewed as a closed loop. In \cite{SV99} an architecture is proposed
that uses a repeating polymer with 8 species to implement a system
having four such tapes.  A rather complicated scheme for
implementing the cooling algorithm is described for this four-tape
machine.

Some versions of the cooling algorithm (\cite{BMRVV}, \cite{FLMR04})
use (classical reversible) 3-bit operations: generalized Toffoli
gates (from which controlled-\SWAP operations can be
implemented).\footnote{By ``generalized Toffoli'' I mean any 3-bit
gate that applies a $\NOT$ operation to one of the bits conditioned
on a specific pattern of the basis states of the other two bits.}
Without access to ancillary bits, the Toffoli cannot be implemented
by classical 2-bit gates (\CNOT and \NOT gates). It can be
implemented without ancilla {\it if} we also have access to
arbitrary single-qubit quantum gates \cite{BBC+95}.  So to implement
the algorithms of \cite{BMRVV}, and \cite{FLMR04} using operations
$o_1$ and $o_2$ would require inherently quantum operations.  An
error analysis of the cooling algorithms is greatly simplified if we
assume it has a ``classical'' implementation, however. Fortunately,
$ABC$-chains naturally support generalized Toffoli operations
directly, since the transition frequency of one species will be
affected by the states of the neighbouring bits of two other
species.

It is worth revisiting the idea put forth in \cite{SV98}, to use an
$ABC$-chain. I propose an alternative set of operations that should
be supported (these are sufficient for cooling, although obviously
not for quantum computing):

\begin{enumerate}
\item[$o^\prime_1$)] Move any
three bits into adjacent positions under a fixed ``tape head''
(which covers three bits).
\item[$o^\prime_2$)] Apply any generalized Toffoli or \CNOT operation to the bits under the tape head.
\end{enumerate}

Using the scheme described in Appendix \ref{append_abc_ring_scheme},
$o^\prime_1$ and $o^\prime_2$ can be implemented on an $ABC$-chain
which is configured as a closed loop.\footnote{We could
alternatively use a linear configuration, but would then have to be
careful about the behaviour at the ends of the chain. One approach
would be to have the chain be long enough so that the bits of
interest are sufficiently far into the interior of the chain that
the effects the ends are irrelevant.} An atom of a fourth type, $D$
is positioned adjacent to some $ABC$-triple selected (arbitrarily)
to be the position of the tape head.

The cooling algorithms work by moving some bits under the tape head
and applying a basic (2-bit or 3-bit) RPC step. The resulting cooler
bits are then moved to one side of the array (tape), while the
hotter bits are moved to the other side. The RPC step is repeated to
cool several bits, and then recursively applied to these cooled
bits.

\section{The Reversible Polarization Compression Step}

We will assume that our initial configuration is some string of
bits, each of which is (independently) in state 0 with some
probability $p>0$.  Equivalently, we assume the bits all have an
identical bias $\B>0$ before applying the polarization compression
step.  The assumption of independence (i.e. a binomial distribution
on the strings) is required for the analysis. \footnote{In
\cite{SV99} it is suggested that by performing an initial
permutation of the bits we can limit our reliance on the assumption
of independence.} Algorithmic cooling only amplifies an existing
bias and hence the initial bias $\B$ must be positive.

The basic idea behind RPC is to implement a permutation that maps
strings with low Hamming weight (i.e. having many 0's) to strings
having a long prefix of 0's.  Because it will be useful to implement
cooling algorithms on systems for which we don't have arbitrary
local control, we will construct RPC permutations based on basic
``RPC steps''.  An RPC step will be a permutation on the states of a
small number of bits (2 or 3 in the examples I consider).  The
overall system will be cooled by recursively applying the basic RPC
step to all the bits.  If we apply the RPC steps to disjoint pairs
or triples of bits at each stage, the assumption of independence
will hold throughout.

In the following sections we will examine candidates for the RPC
step, and discuss how they may be implemented.

\subsection{The 2-bit RPC step}\label{sec_2bit_step}

The algorithms described in \cite{SV99} and \cite{BMRVV} both use a
very simple 2-bit operation for the basic RPC step. The operation
begins with a \CNOT gate.  Suppose the \CNOT is applied to two bits
initially having some positive bias $\B$. After the \CNOT, the
target bit is 0 if both bits were originally equal, and is 1 if both
bits were originally different. In the case that they were both the
same, the control bit has an amplified bias after the \CNOT. So,
conditioned on the outcome of the target bit, the control bit is
either accepted as a new bit with higher bias and is subsequently
moved to the ``colder'' side of the array with a sequence of
controlled-\SWAP operations, or it is rejected and subsequently
moved to the ``warmer'' side of the array.  For specificity, I will
refer to this 2-bit RPC step as ``2BC''.

Suppose the values of the control and target bits before the \CNOT
are $b_c$ and $b_t$ respectively.  Then after the \CNOT the value of
the target bits is $b_c+b_t$.  The control bit is accepted iff this
value equals 0.  The probability that $b_c=0$ given that $b_c+b_t=0$
is
\begin{align}
&\frac{P(b_c=0 \wedge b_t=0)}{P(b_c+b_t=0)}\\
&=\frac{1}{2}+\frac{2\B}{1+\B^2}
\end{align}
and so in this case the bias of the control bit is
\begin{equation}
\B^\prime=\frac{2\B}{1+\B^2}.
\end{equation}
The probability that the control bit is accepted equals the
probability that $b_c+b_t=0$, which is
\begin{equation}
\frac{1+\B^2}{2}.
\end{equation}

If the control bit is rejected, it has bias 0.  To achieve the
polarization compression, the \CNOT must be followed by an operation
that selects the accepted bits to be retained.  This is accomplished
in the 2BC operation by controlled-\SWAP operations that move the
bit to the left or right according to whether it was accepted or
rejected.\footnote{In Section \ref{sec_2BC_3BC_equivalence} we will
show that the \CNOT followed by a controlled-\SWAP actually computes
the majority of three bits, and thus the 2BC operation is equivalent
to the 3BC operation defined in Section \ref{sec_3bit_step}.}

A cooling algorithm can work by recursive application of the 2BC
step across many bits having an initial bias $\B_\init$.  First some
of the bits will be cooled by one application of 2BC, while others
are warmed. The cooled bits will be moved away from the warmed bits,
and then cooled further by another application of 2BC, and so on.
The total number of starting bits required is determined by the
depth of recursion required to obtain a single bit cooled to the
desired target bias.

\subsection{A 3-bit RPC step}\label{sec_3bit_step}

\begin{figure}[h]
\begin{center}
\epsfig{file=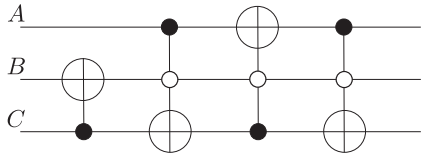, height=2cm}\caption{\small{A circuit for
the 3BC step using \CNOT gates and generalized Toffoli
gates.}}\label{fig_cswap_cooling}
\end{center}
\end{figure}

The algorithm described in \cite{FLMR04} uses a 3-bit reversible
polarization compression step (3BC).  This RPC step is implemented
by a permutation on the basis states of a 3-bit register that has
the effect of increasing the bias of the one of the bits, while
decreasing the bias of the other two.  Experimental demonstration of
the 3-bit RPC step has been conducted using NMR \cite{BMRNL}. The
implementation of the 3BC operation given in \cite{FLMR04} uses a
\CNOT gate followed by a controlled-\SWAP gate. Recall from our
discussion in Section \ref{sec_architecture} that we are assuming
that the bits have already been moved onto an $ABC$-triple under the
``tape-head'', and that we can implement any reversible 3-bit
(classical) operation on them.  The quantum circuit model is a
convenient paradigm for describing the operations\footnote{Current
NMR experiments in algorithmic cooling \cite{BMRNL} do not implement
the 3-bit permutation through a decomposition into a sequence of
gates such as we consider here, but rather use a more direct method.
This direct method is not scalable in the number of bits over which
the majority is being computed.}. Note that the controlled-\SWAP can
be implemented by generalized Toffoli operations, as shown in Figure
\ref{fig_cswap_cooling}. (Approaches for implementing such
generalized Toffoli gates on $ABC$-chains are described for example
in \cite{Llo93} and \cite{Ben00}.)

The permutation implemented by the circuit in Figure
\ref{fig_cswap_cooling} results in the majority value of the three
bits (before the operation) being encoded into bit $A$.  Since we
are only interested in the final bias of bit $A$, we can use any
permutation that has this effect.  In fact, the following claim says
that such a permutation is the best choice for a 3-bit RPC step.

\begin{claim}
Suppose we have a register of $n$ bits independently having
identical bias $\B>0$, where $n$ is odd.  Suppose we want to
implement a permutation that has the effect of increasing the bias
of the first bit as much as possible.  Then the best choice is a
permutation that computes the majority value of the $n$ bits into
the first bit.
\end{claim}

\begin{quote}
{\bf Proof} Since each bit has bias $\B>0$, each bit is
independently 0 with probability $p>\frac{1}{2}$.  An optimal
permutation for increasing the bias of the first bit will be one
which maps the $\frac{2^n}{2}$ most-likely strings to strings having
a 0 in the first bit.  The $\frac{2^n}{2}$ most-likely strings are
precisely those having at least $\ceil{\frac{n}{2}}$ bits in the
state 0.\hspace{5mm}$\square$
\end{quote}

\begin{figure}[h]
\begin{center}
\epsfig{file=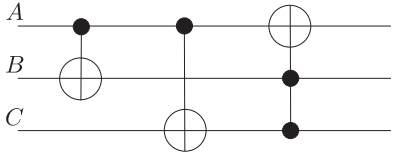, height=2cm}\caption[An alternative circuit
for the cooling step]{\small{An alternative circuit for computing
the majority of three bits can be used for the 3BC
operation.}}\label{fig_3bitmaj}
\end{center}
\end{figure}

The circuit is shown in Figure \ref{fig_3bitmaj} is an alternative
implementation of the 3-bit majority, which is simpler in terms of
Toffoli and \CNOT operations. I will henceforth refer to the
operation implemented by this circuit as 3BC.  Note that the circuit
of Figure \ref{fig_3bitmaj} implements a different permutation that
that implemented by the circuit of Figure \ref{fig_cswap_cooling},
but the effect on bit $A$ (i.e. after tracing-out bits $B$ and $C$)
is the same for both circuits (assuming the input bits are
independently distributed).

Since Toffoli and \CNOT operations are ``classical'' in the sense
that they do not generate nontrivial superpositions given basis
states as inputs, we can analyze the behaviour of the 3BC circuit
entirely in the computational basis.  In the following, I will
restrict the analysis in terms of classical bits.

Consider the effect on the bias of bit $A$ after applying the
circuit of Figure \ref{fig_3bitmaj}. The majority value is computed
into bit $A$. Suppose initially the bias of each of the three bits
is $\B$. So the probability for each bit equaling 0 is initially
$(1+\B)/2$. After the 3BC operation, the probability that bit $A$
(which now equals the majority of the initial values of $A,B,C$)
equals zero is
\begin{align}
p^{(A)}&=\left(\frac{1+\B}{2}\right)^3+3\left(\frac{1+\B}{2}\right)^2\left(\frac{1-\B}{2}\right)\\
&=\frac{1}{4}(2+3\B-\B^3).
\end{align}
So the bias of bit $A$ after the 3BC operation is
\begin{align}
\B^\prime&=2p^{(A)}-1\\
&=\frac{3}{2}\B-\frac{1}{2}\B^3\label{3BC_newB}.
\end{align}

\subsection{Equivalence between the 2BC and 3BC operations}\label{sec_2BC_3BC_equivalence}

Recall that 2BC is \CNOT followed by controlled-\SWAP operations
which moves the control bit (of the \CNOT) to the left or right
conditioned on the state of the target bit. The \CNOT itself has no
effect on the bias of the control bit.  It is the value of the
target bit after the \CNOT that provides some information about the
state of the control bit. In the case that the target bit equals
zero, the control bit is more likely to be 0, and hence has a
greater bias. So the 2BC step is really a method for gaining some
information about which bits are more likely to be zero, and moving
these off to one side.  After a single application of 2BC on two
bits having equal bias, we may or may not be left with a bit having
increased bias.

The 3BC step, on the other hand, deterministically increases the
bias of the third bit at the expense of decreasing the bias of the
other two.  Every time we apply the 3BC step to three bits having
equal bias we are certain to be left with a bit whose bias has been
increased.  This property makes it somewhat simpler to analyze the
efficiency of algorithms based 3BC.  The analysis of the 2BC-based
heat-bath algorithm in \cite{BMRVV} relies on the law of large
numbers, and gives a worse bound than does the analysis of the
3BC-based\par algorithm of \cite{FLMR04}.

In the algorithms of \cite{SV99} and \cite{BMRVV} the \CNOT of the
2BC step is always followed by a controlled-\SWAP operation.  An
important observation is that the \CNOT followed by a
controlled-\SWAP actually computes the three-bit majority (indeed
this is the way the 3BC step was implemented in \cite{FLMR04}).
Specifically, suppose we first apply a \CNOT between bits in states
$b_1$ and $b_2$ (with $b_1$ as the control bit), and then apply a
controlled-\SWAP between $b_1$ and a third bit in state $c$,
controlled on the target bit of the \CNOT being 0.  The final state
of $c$ is
\begin{equation}
b_1c+b_2c+b_1b_2
\end{equation}
which is the majority of $b_1,b_2,c$. So if we explicitly include
the extra target bit of the controlled-\SWAP operation, the 2BC step
is is equivalent to the 3BC step.

This suggests an equivalence between the early algorithms described
in terms of a 2BC operation and algorithms phrased in terms of a
3-bit majority vote (3BC). For this reason, in the following I will
restrict attention to algorithms based on the 3BC operation.

\section{Efficiency}\label{sec_efficiency}

\subsection{The simple recursive algorithm}

We will analyze the efficiency of a simple algorithm that
recursively partitions the string of bits into triplets and applies
3BC to these triplets. After each 3BC step (say on bits $A,B,C$),
the $B$ and $C$ bits which become heated are discarded. Thus at each
level of recursion the total number of bits is reduced by a factor
of 3, and the remaining bits' bias is increased from $\B$ to a new
value
\begin{equation}
\B^\prime=\frac{3}{2}\B-\frac{1}{2}\B^3.
\end{equation}
To simplify the analysis we will approximate $\B^\prime$ by
\begin{equation}
\B^\prime\approx\frac{3}{2}\B.
\end{equation}

After $k$ levels of recursion the bias is increased to
\begin{equation}
\B_k\approx\left(\frac{3}{2}\right)^k \B.
\end{equation}
This gives us an estimate on the number of levels of recursion $k$
required to achieve some target bias $\B_t<1$ on a single bit.
\begin{equation}
k\approx\log_{3/2}\left(\frac{\B_t}{\B}\right).
\end{equation}
Therefore the total number of bits starting at bias $\B$ required to
obtain one bit with a target bias of $\B_t$ is $3^k$ which is
polynomial in $\B_t$.  For example, suppose we start with a bias of
$\B=10^{-5}$ (see \cite{M05}).  Then the number of bits required to
yield a single bit with bias $0.1$ is about $6.9\times 10^{10}$, and
the number required to yield a bit with bias $0.9999$ is about
$3.5\times 10^{13}$.

Note that this analysis has only given the number of bits required.
To obtain a good estimate of the time complexity, we would have to
specify the computational model more precisely, and account for the
time required to shuttle the states around as required by the
architecture and the algorithm.

\subsection{Algorithms using a heat bath}

There are many ways in which the recursive algorithm might be
modified to take advantage of a heat bath.  The heat bath is a
mechanism by which a heated bit can be exchanged for a fresh bit
having initial bias $\B_\init$ (taken from the environment).  For a
rough analysis, we ignore the details of how the heat-bath contact
will be implemented, and assume we can apply an operation which
resets a bit's bias to $\B_\init$ on-demand (this may be an
unrealistically optimistic assumption).

One approach to using the heat bath in a 3BC algorithm is as
follows.  First apply the 3BC step as in the simple recursive
algorithm.  At this point we have $n/3$ bits cooled to $\B^\prime$.
Now, instead of discarding the $2n/3$ bits that were heated in this
process, send them to the heat bath to return them to bias
$\B_\init$. Then partition these $2n/3$ bits into triples, and apply
the 3BC step to them.  This yields another $2n/9$ bits of bias
$\B^\prime$.  Repeat this process until there are fewer than $3$
bits left having bias less than $\B^\prime$ (there will always be
exactly 2 bits left over).  Now we have $n-2$ bits cooled to bias
$\B^\prime$ and we can proceed to the next level of recursion.  As
before, the number of levels of recursion $k$ required to achieve a
bit having some target bias $\B_t<1$ is
\begin{equation}
k\approx\log_{3/2}\left(\frac{\B_t}{\B_\init}\right).
\end{equation}
This time, however, a logarithmic amount additional work is done for
each level of recursion.  By taking this extra time, we save on the
total number of bits required.  After each level of recursion an
additional 2 bits are discarded.  So the total number of bits
required to obtain one bit cooled to $\B_t$ by this method is $2k$
which is polylogarithmic in $\B_t$.  As before, supposing we start
with a bias of $\B_\init=10^{-5}$, then the number of bits required
to yield a single bit with bias $0.1$ is about $46$, and the number
required to yield a bit with bias $0.9999$ is about $57$.

Another approach to using the heat bath is described in
\cite{SMW07}.  Their algorithm repeatedly applies the 3BC step to
three bits having bias values $\B_{j-2},\B_{j-1}$ and $\B_j$.  This
requires more careful analysis.  Consider applying 3BC to three bits
$b_{j-2},b_{j-1},b_j$ having initial bias values $\B_{j-2},\B_{j-1}$
and $\B_j$ respectively, where the majority is computed into the
third bit $b_j$.  The resulting bias of the third bit is
\begin{equation}
\B_j^\prime=\frac{\B_{j-2}+\B_{j-1}+\B_j-\B_{j-2}\B_{j-1}\B_j}{2}.
\end{equation}
Now suppose the first two  bits are sent to the heat bath, and then
run back through the cooling procedure to regain bias values of
$\B_{j-2}$ and $\B_{j-1}$.  Then 3BC is applied again (on the same
three bits, except this time the third bit starts with bias
$\B_j^\prime$). If this process is repeated several times, the bias
of the third bit reaches a steady state value of
\begin{equation}
\frac{\B_{j-2}+\B_{j-1}}{1+\B_{j-2}\B_{j-1}}.
\end{equation}

The algorithm described by \cite{SMW07} is based on this process.
Suppose the algorithm has built-up an array of $k>3$ cooled bits
$b_1,b_2\ldots,b_k$ having bias values $\B_1,\B_2,\ldots,\B_{k}$ in
that order, where
$\B_j=\frac{\B_{j-2}+\B_{j-1}}{1+\B_{j-2}\B_{j-1}}$ for each
$3<j<k$.  Then in the next stage of the algorithm a new bit
$b_{k+1}$ is introduced having the heat-bath bias $\B_0$.  The 3BC
procedure is applied to the three bits $b_{k-1},b_k,b_{k+1}$
repeatedly, where between each application the algorithm is
recursively repeated to re-establish the bias values $\B_{k-1},\B_k$
on bits $b_k,b_{k+1}$.  Repeating this process several times the
bias of bit $b_{k+1}$ will reach the steady state value
$\B_{k+1}=\frac{\B_{k-1}+\B_{k}}{1+\B_{k-1}\B_{k}}$.

Starting with $n$ bits of bias $\B_\init$, the algorithm of
\cite{SMW07} achieves one bit of bias approximately $\B_n=\B_\init
F(n)$, where $F(n)$ is the $n^{th}$ Fibonacci number. This is even
better than the simple recursive heat-bath method described
previously. Starting with a bias of $\B_\init=10^{-5}$, the number
of bits required for this method to yield a single bit with bias
$0.1$ is about $20$, and the number required to yield a bit with
bias $0.9999$ is about $28$. There is a polynomial cost in time
incurred by the repeated re-cooling of bits from the point of
heat-bath contact at the left end of the chain up to $\B_{j-2}$ and
$\B_{j-1}$.

Notice that in the heat bath algorithms we have described, after a
3BC operation the two bits that have become heated by this operation
are both sent to the heat bath.  In the early stages of an
algorithm, this would be sensible, because the 3BC operation will
have warmed those two bits to bias values less than the initial bias
$\B_\init$.  Towards the end of the algorithm, however, 3BC will be
applied to triples of bits that are all very cold bits, and the bits
that become heated may still have bias considerably higher than
$\B_\init$.  In this case, sending these bits to the heat bath does
not seem like the right thing to do.  To analyze the performance of
algorithms, however, it is extremely convenient to assume we always
do so.  If we do not send the two heated bits back to the heat bath
after a 3BC application, the bits' values are no longer described by
independent probability distributions, and bias values are no longer
well-defined.  It is convenient to model the process of a 3BC
application followed by sending the two heated bits to the heat bath
as a single operation, as follows.

\begin{definition}
Consider three bits $b_1,b_2,b_3$ having bias values
$\B_1\leq\B_2\leq\B_3$ respectively. Define $\TBC_\hb$ as the
three-bit majority on $b_1,b_2,b_3$ (where the majority is computed
into $b_3$) followed by sending $b_1$ and $b_2$ to the heat bath.
The bias values of the three bits after this operation are
$\B_\init,\B_\init,\frac{\B_1+\B_2+\B_3-\B_1\B_2\B_3}{2}$
respectively.
\end{definition}

The heat-bath algorithms described above can all be described as a
sequence of operations $(\TBC_\hb, P_1, \TBC_\hb, P_2, \TBC_\hb,
P_3, \ldots)$ where each $\TBC_\hb$ is applied to three bits in some
specific positions (e.g. under a ``tape head''), and each $P_i$ is
some permutations of the positions of the bits in the string. The
following claim shows that the algorithm of \cite{SMW07} is the best
such algorithm (this is not claimed in \cite{SMW07}).

\begin{claim}
Consider a string of bits each having initial bias value $\B_\init$.
Let $\mathcal{A}$ be any cooling algorithm described by a sequence
of operations
\[\TBC_\hb, P_1, \TBC_\hb, P_2, \TBC_\hb, P_3, \ldots\]
where each $\TBC_\hb$ is applied to three bits in some specific
positions (e.g. under a ``tape head''), and each $P_i$ is some
permutation of the positions of the bits in the string.  At any
stage of the algorithm, suppose we arrange the bits in a
nondecreasing order of their bias values $\B_1,\ldots\B_N$.  Then we
have $B_j\leq \B_\init F_j$ for all $1\leq j\leq N$, where $F_j$ is
the $j^\text{th}$ Fibonacci number.
\end{claim}

\begin{quote}
The proof is by induction.  The claim is initially true (before
starting the algorithm) by assumption.  Since the only operation
allowed in $\mathcal{A}$ that changes the bias values is the
$\TBC_\hb$ operation, it suffices to show that after an arbitrary
$\TBC_\hb$ operation the claim is still true. Suppose the ordered
bias values before the 3BC operation are
\[\B_1,\B_2,\ldots, \B_N.\]
Then suppose the $\TBC_\hb$ operation is applied to any three bits,
suppose those having bias values $\B_k,\B_l$ and $\B_m$, where
$k<l<m$. We assume that after the 3BC operation the value of $\B_m$
is not decreased. This is a safe assumption, because otherwise
algorithm $\mathcal{A}$ would have done just as well not to apply
that 3BC operation.

After the $\TBC_\hb$ operation, the new bias of the bit originally
indexed by $m$ is
\[\B_r^\prime =\frac{\B_k+\B_l+\B_m-\B_k\B_l\B_m }{2}.\]
By assumption we have $\B_m\leq \B_\init F_{m}$, $\B_l\leq \B_\init
F_{m-1}$ and $\B_k\leq \B_\init F_{m-2}$.  Since
$F_m=F_{m-1}+F_{m-2}$ by definition, we have
\begin{equation}\label{eqn_bound_Br}
\B_r^\prime\leq \B_\init F_m.
\end{equation}
Now, suppose the re-ordered bias values are
\[\B^\prime_1,\ldots,\B^\prime_N.\]
Since two of the bits were subjected to heat bath contact, we have
$\B^\prime_1=\B^\prime_2=\B_\init$ and $B_j^\prime<\B_j$ for $3\leq
j\leq m-1$.  So the claim is true for the first $m-1$ bias values.
By the ordering we have $\B^\prime_m,\ldots,\B^\prime_{r-1}\leq
\B^\prime_r$, and by (\ref{eqn_bound_Br}) we know these are all at
most $F_m$, so the claim is true for these bias values.  For the
remaining bits we have $\B^\prime_j=\B_i\leq F_j$ for $r+1\leq j\leq
N$, and so the claim is true for them as well.  This completes the
proof.$\qed$
\end{quote}

\subsection{Accounting for the heat bath as a computational
resource}\label{sec_heat_bath_resource}

The heat bath is typically modeled by a process whereby a hot bit is
magically replaced by a fresh bit having the initial bias
$\B_\init$. Usually we would make some assumption about where the
heat-bath contact occurs, for example requiring that only the bit on
the end of a chain can be replaced with a fresh bit.

From a complexity theory point of view, the heat bath is a resource
that should be account for. For modeling the physics of the
situation it might be very convenient to draw a conceptual boundary
between the system we are trying to cool and the heat bath, which
for all practical purposes might be extremely large. Continuing our
previous analogy between heat-bath cooling and a kitchen
refrigerator, if we put the refrigerator in a large enough room we
won't have to account for the fact that the room itself is gradually
heated by the radiator on the back of the fridge.  While heat-bath
techniques appear to drastically reduce the number of bits required
to achieve a target bias, it should be recognized that this hasn't
come for free.  The extra bits ultimately come from the heat bath.
In practice, it may be very reasonable to assume we get these bits
``for free'', since we don't have to exercise control over the heat
bath the way we do with the bits directly involved in the algorithm.

\section{Accounting for errors in an analysis of cooling}

In the following sections I investigate the performance of cooling
algorithms when errors can occur in the RPC step. The bounds I will
derive will apply to cooling algorithms that are based on recursive
application of the 3BC step, where the step is always applied to 3
bits that have been previously cooled to equal bias values. In
Section \ref{sec_general_3BC_algs} I discuss how the same approach
can be applied to analyze more general algorithms based on the 3BC
step. I do not account for errors that might occur between
applications of the RPC steps, such as when bits are being shuttled
around, or placed in an external heat bath. For this reason the
bounds apply quite generally, independent of implementation details
and low-level algorithmic details.

The most general way to analyze the effect of errors on a quantum
circuit is to examine the effect of the errors on the density matrix
of the state as it evolves through the circuit.  As we observed
above, the 3BC step can be implemented by classical operations, and
can be analyzed entirely in the computational basis. I therefore
perform the error analysis in a classical setting.

Suppose we implement the RPC operation in a system subject to errors
described by a set of error patterns $\{S_j\}$.  The error pattern
is a record of what errors actually occurred. For each error pattern
$S_j$ we can analyze the effect by considering a new circuit
containing the original RPC circuit as well as the error operations
that occurred.  We can then find the probability $p_j$ that the
cooled bit would be in state 0 after applying this new circuit. Thus
the probability that the cooled bit equals 0 for the overall process
is
\begin{equation}
p=\sum_j p_j\Pr(S_j)
\end{equation}
where $\Pr(S_j)$ is the probability that error pattern $S_j$ occurs.
The new bias of the cooled bit after the process is then
\begin{equation}
\B^\prime=2p-1
\end{equation}
Equivalently, we could compute the new bias $\B_j^\prime$ of the
cooled bit resulting from application of the RPC step for each error
pattern $S_j$, and take a weighted sum of these bias values
(weighted by the probabilities $\Pr(S_j)$).
\begin{equation}
\B^\prime=\sum_j \B_j^\prime\Pr(S_j).
\end{equation}
After obtaining the new bias $\B^\prime$ of the cooled bit after the
overall process, we can obtain theoretical limits on the performance
of the cooling algorithm by analyzing the condition
\begin{equation}
\B^\prime > \B
\end{equation}
where $\B$ is the bias before the RPC and error channel were applied
(this simply says that the bias should be greater after application
of the 3BC step). In practice, to analyze the inequality
\begin{equation}\label{general_bias_boost_inequal}
\B^\prime - \B > 0
\end{equation}
we study the expression $\B^\prime - \B$, which for the error models
we consider will be a quadratic or cubic polynomial in $\B$ (and
also a function of the error rates).  By studying the roots of this
polynomial we can find ranges of values for the error rates for
which inequality (\ref{general_bias_boost_inequal}) has solutions
$\B>0$, and also obtain the maximum value of $\B$ which is a
solution (this maximum value will be the maximum bias achievable by
the RPC step for the given error rates).

\section{The symmetric bit-flip channel}

The first error model we will consider is the symmetric bit-flip
model, in which a bit's value is flipped with probability
$\eps<\frac{1}{2}$ (``symmetric'' in this context means that the
probability of a bit flip is independent of the initial state of the
bit).

If the bit-flip channel is applied to a bit initially having bias
$\B$, the result is a bit with bias $-\B$.

\subsection{3BC followed by a symmetric bit-flip error}\label{sec_sym_after_3BC}

We will now consider the case in which a bit-flip error can occur
after the 3BC step has been performed (and errors do not occur
between application of the gates in Figure \ref{fig_3bitmaj}).

There are two error patterns. Pattern $S_1$ represents the case
where a bit flip does not occur. In this case, the final bias of bit
$A$ is
\begin{equation}
\B_1^{\prime}=\frac{3}{2}\B-\frac{1}{2}\B^3
\end{equation}
as we found in Section \ref{sec_3bit_step} (equation
(\ref{3BC_newB})). The error pattern $S_2$ represents the case where
the bit flip occurs on the newly biased bit.  In this case, the bias
is negated, and so the new bias is
\begin{equation}
\B_2^{\prime}=-\frac{3}{2}\B+\frac{1}{2}\B^3
\end{equation}
So the new bias of $A$ for the overall process is
\begin{align}
\B^\prime&=(1-\eps)\B_1^\prime+\eps \B_2^\prime\\
&=\left(\frac{3}{2}\B-\frac{1}{2}\B^3\right)(1-2\eps)\label{sym_after_3BC_newB}.
\end{align}

Then the condition that $\B^\prime > \B$  gives
\begin{equation}\label{sym_after_3BC_thresh_inequal}
-(1-2\eps)\B^2-6\eps+1 > 0
\end{equation}
which leads to
\begin{align}
\eps& < \frac{1}{2}-\frac{1}{3-\B^2}\\
& < \frac{1}{6}.\label{sym_after_2BC_thresh}
\end{align}
So for this simple error model $\eps_{\th}=1/6$ is an error
threshold beyond which the 3BC procedure can have no positive effect
on the bias (regardless of how low the initial bias is).  For a
fixed error rate $\eps<\eps_\th$ a bound on the maximum bias that
will be achievable is obtained by solving for $\B$ in
(\ref{sym_after_3BC_thresh_inequal})):
\begin{equation}\label{sym_after_3BC_Blim}
\B < \sqrt{\frac{1-6\eps}{1-2\eps}}=\B_\lim.
\end{equation}
Approximating the expression to second-order gives
\begin{equation}\label{sym_after_3BC_Blim_approx}
\B_\lim\approx 1-2\eps-6\eps^2.
\end{equation}

Once the bias exceeds $\B_\lim$, the 3BC procedure will no longer be
effective in increasing the bias, and the algorithm will yield no
further improvement.  So $\B_\lim$ represents the limit of the bias
that can be achieved by any cooling algorithm that is based on the
3BC step, under this error model.

For error rates $\eps$ below 1\%, the approximate value in
(\ref{sym_after_3BC_Blim_approx}) is within 0.01\% of the value in
(\ref{sym_after_3BC_Blim}).

\subsection{Symmetric bit-flip errors during application of
3BC}\label{sec_sym_during_3BC}

We will now do a more careful analysis accounting for the
possibility of errors occurring during the application of the 3BC
step. Specifically, we consider independent bit-flip errors on each
bit with probability $\eps$, where the errors can occur at each time
step; that is, immediately after the application of any gate in the
circuit of Figure \ref{fig_3bitmaj} (equivalently after the
application of each $o_2^\prime$ operation).  This is only one
possible decomposition of the majority-vote operation into a
sequence of basic operations, but it serves to illustrate the
technique for analysis. A similar analysis can easily be conducted
given an alternative decomposition of the majority vote into a
sequence of basic operations.

\begin{figure}[h]
\begin{center}
\epsfig{file=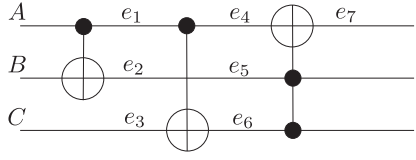, height=2cm}\caption[The majority
circuit with relevant error positions shown.]{\small{The majority
circuit with relevant error positions shown. The binary variables
$e_i$ are taken to have the value 1 if a bit-flip error occurred in
the relevant location (otherwise $e_i=0$).}}\label{fig_3bitmaj_err1}
\end{center}
\end{figure}

There are 9 possible sites for bit-flip errors in Figure
\ref{fig_3bitmaj}, but two of these can be ignored (errors on the
$B$ or $C$ bits after the final Toffoli have no effect on the final
bias of the $A$ bit). Figure \ref{fig_3bitmaj_err1} illustrates the
circuit including the possible error operations. The binary
variables $e_i$ shown on the circuit are taken to be ``1'' if a
bit-flip error occurs in that location, and ``0'' otherwise.

Suppose the value of the $(A,B,C)$ register is initially $(a,b,c)$,
where $a$, $b$, and $c$ are the binary values of the three bits.
Analyzing the circuit in Figure \ref{fig_3bitmaj_err1}, the final
value of the $A$ bit is found to be
\begin{equation}\label{3bitmaj_err1_newA}
a+e_1+e_4+e_7+(a+b+e_2+e_5)(a+c+e_1+e_3+e_6)\mod 2.
\end{equation}
Since the errors occur independently with probability $\eps$ at each
position, the probability associated with each error pattern
$S_i=(e_1,e_2,\ldots,e_7)$ (where $i=\sum_{k=0}^6e_k2^k$ indexes the
possible patterns) can be evaluated as
\begin{equation}\label{3bitmaj_err1_prob_pattern}
\text{Pr}(S_i)=\eps^{e_1+e_2+e_3+e_4+e_5+e_6+e_7}
(1-\eps)^{\bar{e}_1+\bar{e}_2+\bar{e}_3+\bar{e}_4+\bar{e}_5+\bar{e}_6+\bar{e}_7}
\end{equation}
where $\bar{x}\equiv 1+x\mod 2$. Initially, the probability that
each bit $a,b$ or $c$ equals 0 is $p=\frac{B_0+1}{2}$.  So the tuple
$(a,b,c,e_1,\ldots,e_7)$ describes the situation where the register
was initially in the state $(a,b,c)$ and the error described by
$(e_1,e_2,\ldots,e_7)$ occurred, and this happens with probability
\begin{equation}\label{3bitmaj_sym_err_prob_scen}
\text{Pr}(a,b,c,e_1,\ldots,e_7)\equiv(1-p)^{a+b+c}p^{\bar{a}+\bar{b}+\bar{c}}\eps^{e_1+e_2+e_3+e_4+e_5+e_6+e_7}
(1-\eps)^{\bar{e_1}+\bar{e_2}+\bar{e_3}+\bar{e_4}+\bar{e_5}+\bar{e_6}+\bar{e_7}}.
\end{equation}
Let $p^{(A)}$ be the probability that the final value of $A$ for the
overall process equals 0. The value of $p^{(A)}$ is obtained by
adding the probabilities $\text{Pr}(a,b,c,e_1,\ldots,e_7)$ over all
those tuples $(a,b,c,e_1,\ldots,e_7)$ for which the value of
(\ref{3bitmaj_err1_newA}) equals 0.  The new bias of $A$ is then
determined as
\begin{equation}\label{newbias_pA}
\B^\prime=2p^{(A)}-1.
\end{equation}
This value is
\begin{equation}
\B^\prime=(2\eps-1)^3\left[1+4\eps^2(\eps-1)-4p\eps(6\eps^2-8\eps+3)-2p^2(2p-3)(2\eps-1)^3\right]
\end{equation}
which can be expressed in terms of the original bias by substituting
$p=\frac{\B+1}{2}$:
\begin{equation}\label{sym_during_3BC_newB}
\B^\prime=
\frac{1}{2}\B(1-2\eps)^3\left(3-6\eps+4\eps^2-\B^2(1-2\eps)^3\right).
\end{equation}

Now the condition $\B^\prime>\B$ leads to
\begin{equation}\label{sym_during_3BC_inequal}
-2+(1-2\eps)^3\left(3-6\eps+4\eps^2-\B^2(1-2\eps)^3\right)>0.
\end{equation}

The expression on the left side of (\ref{sym_during_3BC_inequal})
represents the improvement in the bias.  It decreases monotonically
as $\B$ increases from 0, and so an upper bound can be obtained by
setting $\B=0$.  Then, by studying the real roots of the resulting
polynomial in $\eps$ we can determine the range of values for which
the improvement is positive.  By numerical computation, the
threshold is found to be

\begin{equation}\label{sym_during_3BC_thresh}
\eps < 0.048592 \equiv\eps_\th.
\end{equation}

For a fixed $\eps < \eps_{\th}$, inequality
(\ref{sym_during_3BC_inequal}) also gives a bound on the maximum
bias achievable by the 3BC step under the given error model.
\begin{equation}\label{sym_during_3BC_Blim}
\B < \frac{\sqrt{1-24\eps+76\eps^2-120\eps^3+96\eps^4-32\eps^5}}
{(1-2\eps)^3}\equiv \B_\lim
\end{equation}

For small values of $\eps$, we can approximate
(\ref{sym_during_3BC_Blim}) to second order.
\begin{equation}\label{sym_during_3BC_Blim_approx}
\B_\lim\approx 1-6\eps-82\eps^2.
\end{equation}

For error rates $\eps$ below 1\%, the approximate value in
(\ref{sym_during_3BC_Blim_approx}) is within 0.1\% of the value in
(\ref{sym_during_3BC_Blim}).

\section{Debiasing errors}\label{sec_debiasing_errors}

We will now consider a more general error model for a classical bit.
Under this error model, called the {\it asymmetric bit-flip
channel}, a bit transforms from 0 to 1 with some probability
$\eps_0$, and transforms from 1 to 0 with some probability $\eps_1$.

A fixed-point probability distribution for the asymmetric bit-flip
channel is
\begin{align}
p[0]&=\frac{\eps_1}{\eps_0+\eps_1}\\
p[1]&=\frac{\eps_0}{\eps_0+\eps_1}.
\end{align}

If left to evolve for under the symmetric channel, a bit will
eventually settle to a bias value of
\begin{equation}
\B_\steady=\frac{\eps_1-\eps_0}{\eps_0+\eps_1}.
\end{equation}
The rate at which the bias approaches this fixed point is related to
$(\eps_0+\eps_1)$.

It will be convenient to make a couple of assumptions about the
error rates.  First, we will assume that errors cause the system to
tend back to the initial bias $\B_\init$ (which would be the same
as, or close to, the bias of the ``heat bath'' for cooling
algorithms that use this device). That is,
\begin{equation}
\B_\steady=\B_\init.
\end{equation}

In other words, errors cause a partial debiasing of the cooled bits
(ideally, this will happen very slowly, and so a the value for the
sum of the error rates, $(\eps_0+\eps_1)$, will be small).   In the
following, I will refer to this type of asymmetric bit-flip error as
a {\it debiasing error}.

Since $\B_\init>0$, we have
\begin{equation}
\eps_1-\eps_0>0.
\end{equation}

We will also assume that the error rates $\eps_0$ and $\eps_1$ are
both less than $\frac{1}{2}$.  In this case we have
\begin{align}
\eps_1-\eps_0<\B_\init.
\end{align}
Since we assumed that the bias of the bit being cooled starts at
$\B_\init$ and is thereafter nondecreasing, we can say that at any
stage of the algorithm we have
\begin{align}
\eps_1-\eps_0<\B
\end{align}
where $\B$ is the current bias of the bits that the RPC step is
being applied to.

Consider what happens to a bit initially having some bias $\B$ when
we apply the asymmetric bit-flip channel once.  A simple calculation
shows the resulting bias to be
\begin{equation}\label{1_asym_step_newB_e0e1}
\B^\prime=\B(1-(\eps_0+\eps_1))+(\eps_1-\eps_0).
\end{equation}
In the following analysis, it will be convenient to make a change of
variables, letting
\begin{align}
s&\equiv \eps_0+\eps_1\text{ , and}\\
d&\equiv \eps_1-\eps_0.
\end{align}
Then our assumptions are $s<1$, and $0<d<\B$, and equation
(\ref{1_asym_step_newB_e0e1}) becomes
\begin{equation}\label{1_asym_step_newB}
\B^\prime=\B(1-s)+d.
\end{equation}
Notice that $d<s$ is also an obvious condition.

Consider the special case of the symmetric bit-flip channel. In this
case $\B_\steady=0$, and so $\B_\steady<\B_\init$. This is why we
obtained positive threshold error rates for the RPC step to increase
the bias. Now, under our assumption $\B_\steady=\B_\init$, we will
not obtain such a threshold.  Even with high error rates (fast
debiasing) the RPC step will increase the bias above $\B_\init$ by
some positive amount.

It is still important to analyze the effect of these errors on the
RPC step, because they will imply a limiting value on the highest
bias achievable.  The RPC step tends to increase the bias away from
the value $\B_\init=d/s$, while the errors tend to force the bias
back towards $B_\init$.  The maximum achievable value of $\B$ will
be determined by $d$ and $s$, or equivalently, by $\B_\init$ and
$s$. Recall that $s$ can be seen as a measure of the rate at which
the errors force the bias towards the initial value $\B_\init$. Thus
the maximum achievable bias is limited by the initial bias, and by
the rate at which errors cause the system to tend back to the
initial bias.

\subsection{3BC followed by a debiasing error}\label{sec_asym_after_3BC}

Consider the scenario in which a debiasing error may occur
immediately after the 3BC operation. The bound obtained here will
apply regardless of how the 3BC step is implemented. Assuming all
three bits initially start with bias $\B$, the bias of bit $A$ after
the process (the 3BC circuit followed by a debiasing error) is
\begin{equation}\label{asym_after_3BC_newB}
\B^\prime=\left(\frac{3}{2}\B-\frac{1}{2}\B^3\right)(1-s)+d.
\end{equation}
The condition that $B^\prime > B$ leads to
\begin{equation}\label{asym_after_3BC_inequal_newB}
\B^3(s-1)+\B(1-3s)+2d > 0.
\end{equation}
For values of $s<1/3$ (recall the threshold condition $\eps<1/6$ we
obtained in Section \ref{sec_sym_after_3BC}) and for $d<s$, the
cubic polynomial on the left-hand side of
(\ref{asym_after_3BC_inequal_newB}) has one positive real root (and
the value of this root will be less than 1).  A positive value of
$\B$ will satisfy inequality (\ref{asym_after_3BC_inequal_newB})
only if it is less than than the value of this root. That is,
\begin{small}
\begin{equation}\label{asym_after_3BC_Blim}
\B < \frac{i\left(-3(\sqrt{3}-i)(s-1)(3s-1)+(\sqrt{3}+i)
(-27d(s-1)^2+\sqrt{729d^2(s-1)^4+(-3+12s-9s^2)^3})^\frac{2}{3}\right)}
{6(s-1)\left(-27d(s-1)^2+\sqrt{729d^2(s-1)^4+(-3+12s-9s^2)^3}\right)^\frac{1}{3}}.
\end{equation}
\end{small}
The appearance of nonreal numbers in (\ref{asym_after_3BC_Blim}) is
unavoidable\footnote{This is {\it Casus Irreducibilis}: in certain
cases, any expression for the roots of a cubic polynomial in terms
of radicals must involve nonreal expressions, even if all the roots
are real.}. To second order in $d$ and $s$,
(\ref{asym_after_3BC_Blim}) gives
\begin{equation}\label{asym_after_3BC_Blim_approx_sd}
\B_\lim \approx1-s+d-\frac{3}{2}s^2-\frac{3}{2}d^2+3ds.
\end{equation}
In the symmetric case, the bound
(\ref{asym_after_3BC_Blim_approx_sd}) agrees with the bound
(\ref{sym_after_3BC_Blim_approx}) which we found in Section
\ref{sec_sym_after_3BC}.

In terms of $s$ and $\B_\init$,
(\ref{asym_after_3BC_Blim_approx_sd}) is
\begin{equation}\label{asym_after_3BC_Blim_approx}
\B_\lim \approx 1-s-\frac{3}{2}s^2+\B_\init s +3\B_\init
s^2-\frac{3}{2}\B_\init^2s^2.
\end{equation}

For error rates less than 1\%, the approximate value
(\ref{asym_after_3BC_Blim_approx}) agrees with the actual value to
within $10^{-5}$.

\subsection{Debiasing errors during application of 3BC}\label{sec_asym_during_3BC}

We will now consider the error model in which debiasing errors can
occur at each time step (i.e. immediately after the application of
any gate in the circuit of Figure \ref{fig_3bitmaj}, or equivalently
after each $o_2^\prime$ operation). The analysis is performed
similarly to what we did in Section \ref{sec_sym_during_3BC}, by
considering the probability associated with each binary tuple
$(a,b,c,e_1,\ldots,e_7)$ for which the resulting value of bit $A$
equals 0.  For the asymmetric model, by tracing through the circuit
of Figure \ref{fig_3bitmaj}, we find that equation
(\ref{3bitmaj_sym_err_prob_scen}) generalizes to
\begin{small}
\begin{equation}\label{3bitmaj_asym_err_prob_scen}
\Pr(a,b,c,e_1,\ldots,e_7)\equiv(1-p)^{a+b+c}p^{\bar{a}+\bar{b}+\bar{c}}
\left(\eps_0^{\sum_{i=1}^7\bar{\phi}_ie_i}\right)
\left((1-\eps_0)^{\sum_{i=1}^7\bar{\phi}_i\bar{e}_i}\right)
\left(\eps_1^{\sum_{i=1}^7\phi_ie_i}\right)\left((1-\eps_1)^{\sum_{i=1}^7\phi_i\bar{e}_i}\right)
\end{equation}
\end{small}
where $\bar{x}\equiv(1+x\mod 2)$ and
\begin{align}
\phi_1&=a\\
\phi_2&=a+b\mod 2\\
\phi_3&=c\\
\phi_4&=\phi_1+e_1\mod 2\\
\phi_5&=\phi_2+e_2\mod 2\\
\phi_6&=\phi_3+e_3+\phi_4\mod 2\\
\phi_7&=\phi_4+e_4+(\phi_5+e_5)(\phi_6+e_6) \mod 2.
\end{align}
Again we can sum the probabilities $\Pr(a,b,c,e_1,\ldots,e_7)$ over
those tuples for which the final value of bit $A$ (given by equation
(\ref{3bitmaj_asym_err_prob_scen}) equals 0, and compute then new
bias.  The new bias, approximated to second-order in $s$ and $d$, is
\begin{equation}\label{asym_during_3BC_newB}
\B^\prime\approx\frac{1}{2}\left((5d+4d^2-6sd)+(3-12s+19s^2-d^2+4ds)\B+d\B^2+(-1+6s-15s^2)\B^3\right).
\end{equation}
Then the condition $\B^\prime>\B$ gives
\begin{equation}\label{asym_during_3BC_inequal_newB}
(5d+4d^2-6sd)+(1-12s+19s^2-d^2+4ds)\B+d\B^2+(-1+6s-15s^2)\B^3>0.
\end{equation}
For values of $s\lesssim 0.04$ (recall the threshold condition we
obtained in Section \ref{sec_sym_during_3BC}) and for $d<s$, the
cubic polynomial on the left-hand side of
(\ref{asym_during_3BC_inequal_newB}) has one positive real root
(whose value will be less than 1, modulo the error in the
second-order approximation). A positive value of $\B$ will satisfy
inequality (\ref{asym_during_3BC_inequal_newB}) only if it is not
greater than the value of this root, which is (to second order in
$s$ and $d$)
\begin{equation}\label{asym_during_3BC_Blim_approx_sd}
\B_\lim\approx 1-3s+3d-9d^2-\frac{41}{2}s^2+32ds.
\end{equation}
In the symmetric case, the bound
(\ref{asym_during_3BC_Blim_approx_sd}) agrees with the bound
(\ref{sym_during_3BC_Blim_approx}) that we obtained in Section
\ref{sec_sym_during_3BC}. In terms of $s$ and $\B_\init$ we have,

\begin{equation}\label{asym_during_3BC_Blim_approx}
\B_\lim\approx 1-3s-\frac{41}{2}s^2+3\B_\init s+32\B_\init
s^2-9\B_\init^2 s^2.
\end{equation}

For error rates less than 1\%, the approximate value
(\ref{asym_after_3BC_Blim_approx}) agrees with the actual value
(\ref{asym_after_3BC_Blim}) to within $10^{-4}$.

\section{More general algorithms based on
3BC}\label{sec_general_3BC_algs}

In all of the above error analysis, we have assumed that the 3BC
step is applied to three bits having identical bias at each stage of
the algorithm.  Recall in Section \ref{sec_efficiency} it was
mentioned that an algorithm proposed in \cite{SMW07} is structured
somewhat differently, and applies the 3BC step to three bits having
different bias values $\B_{j-2}$, $\B_{j-1}$ and $\B_j$. We can
still learn something by performing the previous analysis assuming
all three bits have bias $\text{max}(\B_{j-2},\B_{j-1},\B_j)$, but
it is worth briefly considering how we could analyze this more
general scenario directly.  Consider applying the debiasing error
channel with error parameters $\eps_0$ and $\eps_1$ immediately
after the 3BC step is applied. In this case, the bias of the third
bit after the process is
\begin{equation}\label{equation_general_3BC_alg_err1}
\frac{\B_{j-2}+\B_{j-1}+\B_j-\B_{j-2}\B_{j-1}\B_j }{2}(1-s)+d
\end{equation}
(recall $s=\eps_0+\eps_1$ and $d=\eps_1-\eps_0$). As in Section
\ref{sec_debiasing_errors}, we assume that the error parameters
satisfy $s<1$, $d>0$ and $d$ is less than each of $\B_{j-2}$,
$\B_{j-1}$ and $\B_j$.  We also assume that $\frac{d}{s}$ is less
than each of $\B_{j-2}$, $\B_{j-1}$ and $\B_j$ so that the errors
are indeed tending the system towards a lower bias.

Now suppose we proceed as in \cite{SMW07} and send the first two
bits back to the heat bath, re-cool them up to bias values
$\B_{j-2}$ and $\B_{j-1}$, and again apply 3BC. Without errors, we
mentioned previously that by repeating this process several times
the third bit reaches a steady-state bias value of
\begin{equation}
\frac{\B_{j-2}+\B_{j-1}}{1+\B_{j-2}\B_{j-1}}.
\end{equation}
With the debiasing error channel being applied after every
application of 3BC, this steady-state bias value is reduced to
\begin{equation}\label{equation_general_3BC_alg_err2}
\frac{(\B_{j-2}+\B_{j-1})(1-s)+2d}{1+\B_{j-2}\B_{j-1}(1-s)+s}.
\end{equation}
Equations (\ref{equation_general_3BC_alg_err1}) and
(\ref{equation_general_3BC_alg_err1}) can be used to analyze can be
used to analyze more general algorithms based on repeated
application of 3BC, including the Fibonacci sequence algorithm
proposed in \cite{SMW07}, under the effect of debiasing errors that
may occur after each application. We could similarly decompose the
3BC step into a suitable sequence of discrete operations, and
proceed as we have done above to analyze the effect of errors that
may occur after each discrete step.

\section{Conclusions and other considerations}

I have studied the performance of cooling algorithms that use the
3-bit majority as the compression step (e.g. \cite{FLMR04},
\cite{SMW07}) and argued that previously discovered algorithms (e.g.
\cite{SV99}, \cite{BMRVV}) can be re-cast in this way. I have proven
the optimality of the best such algorithm for obtaining one cold bit
with the fewest possible number of initially mixed bits. An error
analysis of these algorithms has been conducted, first under a
simple error model (symmetric bit-flip), and then under a more
realistic model of debiasing. Since the implementations of the RPC
steps are inherently ``classical'' (states do not leave the
computational basis), it is reasonable to restrict attention to
these classical error models. In each case, I first derived some
bounds assuming that errors may occur immediately after the RPC
step. Since this may be taken as a best-case scenario, then these
bounds apply regardless of the implementation. I also derived bounds
assuming that the 3BC cooling step is implemented by a sequence of
physical operations that simulate a sequence of \CNOT and Toffoli
gates (i.e. a sequence of $O_2^\prime$ operations). Specifically we
considered the simplest such arrangement for implementing the 3BC
step, shown in Figure \ref{fig_3bitmaj}. The results are summarized
below (approximated to second-order).

\vspace{5mm} {\renewcommand\arraystretch{1.5}
\begin{tabular}{*{3}{|l}|}
\hline
{\bf Error Model}& {\bf Threshold} & {\bf Maximum achievable bias}\\
\hline Symmetric bit-flip after 3BC &$\eps<\frac{1}{6}$ & $1-2\eps-6\eps^2$  \\
\hline Symmetric bit-flip during 3BC &$\eps\lesssim 0.048592$&$1-6\eps-82\eps^2$ \\
\hline debiasing error after 3BC &N/A & $1-s-\frac{3}{2}s^2+\B_\init
s +3\B_\init
s^2-\frac{3}{2}\B_\init^2s^2$\\
\hline debiasing error during 3BC &N/A&
$1-3s-\frac{41}{2}s^2+3\B_\init s+32\B_\init
s^2-9\B_\init^2 s^2$ \\
\hline
\end{tabular}}
\vspace{5mm}

Given a specific low-level implementation of a cooling algorithm,
specified as a sequence of pulses applied to an $ABC$-chain or some
other suitable hardware, a detailed error analysis could be
conducted in a manner similar to the approach I have taken here. For
specific cooling algorithms it will also be interesting to analyze
the effects of errors occurring between applications of the RPC step
(for example, while the bits are being permuted to move the required
bits into position for the next application of the cooling step). By
studying the time-complexity of a specific algorithm implemented on
a specific architecture, we can determine the balance between the
rate at which the algorithm increases the bias, and the rate at
which debiasing errors are causing the bias to decrease.

Cooling algorithms can be built from basic steps other than the
3-bit majority. For those that have ``classical'' implementations
(that is, can be built from some sequence of generalized Toffoli
gates) the approach we have taken here could be employed to conduct
a similar error analysis.  For basic RPC steps operating on more
than 3 bits, the kind of analysis we have performed here would
require examining higher-order polynomials, and this may have to be
done numerically.

For RPC steps that are implemented ``quantumly'' (i.e. using gates
that force states to leave the the computational basis), more
general quantum error models will have to be considered, and a
different approach to the error analysis will be required.

\section*{Acknowledgements}
This research was supported by MITACS (Mathematics of Information
Technology and Complex Systems), NSERC (National Science and
Engineering Research Council), CSE (Communications Security
Establishment), CFI (Canadian Foundation for Innovation), ORDCF
(Ontario Research and Development Challenge Fund), and PREA
(Premier's Research Excellence Awards). I would like to particularly
thank Mark Saaltink for the many interesting conversations that made
this work possible.

\appendix
\section{Architecture using an $ABC$-chain}\label{append_abc_ring_scheme}

Recall the following two operations we proposed that should be
supported by an NMR computer for implementing algorithmic cooling.

\begin{enumerate}
\item[$o^\prime_1$)] Move any
three bits into adjacent positions under a fixed ``tape head''
(which covers three bits).
\item[$o^\prime_2$)] Apply any generalized Toffoli or \CNOT operation to the bits under the tape head.
\end{enumerate}

Here we sketch an method for implementing $o_1^\prime$ and
$o_2^\prime$ on an $ABC$-chain which is configured as a closed loop.
Note that we could alternatively use a linear configuration, but
would then have to be careful about the behaviour at the ends of the
chain (one approach would be to have the chain be long enough so
that the bits of interest are sufficiently far into the interior of
the chain that the effects the ends are irrelevant).  We will also
assume that the loop consists of an odd number of $ABC$-triples.

An atom of a fourth type, $D$ is positioned adjacent to some
$ABC$-triple selected (arbitrarily) to be the position of the tape
head.

We assume that the physical system directly supports the
implementation of a generalized Toffoli or \CNOT operation on all
the $ABC$-triples in parallel (or alternatively an all
$BCA$-triples, or on all $CAB$-triples)\footnote{details are
described in \cite{Llo93}.}. We also assume that we can implement
any such operation on \emph{only} the $ABC$-triple under the tape
head, by making use of the effect of the proximity of the $D$ atom.
So operation $o_2$ is directly supported by the hardware. Given
these primitives, we focus on the problem of implementing operation
$o_1$.

For clarity of exposition, we will refer to the physical bits of
species $A,B,C$ as ``cells'' of ``types'' $A,B,C$.  When we talk
about ``moving a bit to a cell'', we are referring to a sequence of
logical operations (usually nearest-neighbour \SWAP operations) that
permute the logical states of the physical bits on the chain.  We
will use the following lemma.

\begin{lemma}
For any pair of bits initially occupying adjacent cells, there
exists a sequence of primitive operations that has the effect of
bringing those bits into adjacent positions under the tape head.
\end{lemma}

\begin{quote}
\textbf{Proof Sketch}  Let $S_{AB}$ be the operation that swaps the
states of the bits on the $A$-cells with the bits in the
neighbouring-$B$ cells. The sequence $(S_{AC},S_{AB},S_{BC},S_{AB})$
has the effect of moving every bit initially in an $A$-cell to the
$A$-cell of the next $ABC$-triple to the left (counterclockwise). It
also moves every bit initially in a $C$-cell to the $C$-cell of the
next $ABC$-triple to the right (clockwise).  It leaves the bits on
the $B$-cells fixed.  By permuting the labels of the species we have
similar sequences for moving the bits in the $A$- and $B$-cells,
while keeping the bits in the $C$-cells fixed. Suppose we have a
pair of bits $(b_i,c_i)$ in adjacent $B$- and $C$-cells, that we
wish to move into adjacent positions under the tape head. First we
apply the sequence that moves the bits in the $A$- and $B$-cells
(keeping the bits in the $C$-cells fixed) until $b_i$ is in the
$B$-cell under the tape-head. Then we apply the sequence that moves
the bits in the $A$- and $C$-cells (keeping the bits in the
$B$-cells fixed), until $c_i$ is in the $C$-cell under the tape head
(beside $b_i$). Similar procedures will bring any pairs of adjacent
bits into adjacent positions under the tape
head.\hspace{4mm}$\square$
\end{quote}

From the Lemma, it follows that we can implement a nearest-neighbour
\SWAP operation between any adjacent pair of bits on the tape. First
we move the pair under the tape-head, and then use an $o_2^\prime$
operation to \SWAP the states of these bits.  Finally use the
sequence operation of $o_2^\prime$ to move all the bits back to
their corresponding original positions (modulo the swapped pair).
Now it follows that we can implement an arbitrary permutation of the
bits on the tape (by a suitable sequence of nearest-neighbour
transpositions), of which operation $o_1^\prime$ is a special case.
Notice that this also allows us to perform the permutations of the
bits required to move cooled bits to one side of the tape and move
warmer bits to the other side, as would be required between
applications of the basic cooling step.

\end{document}